\documentclass[prl,aps,showpacs,twocolumn,superscriptaddress,amsmath,amssymb,floatfix]{revtex4}
\usepackage{graphicx}
\usepackage{color}

\begin{document}
\title{Vacuum degeneracy of a circuit-QED system in the ultrastrong coupling regime}

\author{Pierre Nataf}
\author{Cristiano Ciuti}
\email[E-mail: ]{cristiano.ciuti@univ-paris-diderot.fr}
\affiliation{Laboratoire Mat\'eriaux et Ph\'enom\`enes Quantiques,
Universit\'e Paris Diderot-Paris 7 and CNRS, \\ B\^atiment Condorcet, 10 rue
Alice Domont et L\'eonie Duquet, 75205 Paris Cedex 13, France}
\affiliation{}
\begin{abstract}
We investigate theoretically the quantum vacuum properties of a chain of $N$ superconducting Josephson atoms inductively coupled to a transmission line resonator.  We derive the quantum field Hamiltonian for such circuit-QED system, showing that, due to the type and strength of the interaction, a quantum phase transition can occur with a twice degenerate quantum vacuum above a critical coupling. In the finite-size case, the degeneracy is lifted, with an energy splitting decreasing exponentially with  increasing values of $g^2 N^2$, where $g$ is the dimensionless vacuum Rabi coupling per artificial atom.  We determine analytically the ultrastrong coupling  asymptotic expression of the two degenerate vacua for an arbitrary number of artificial atoms and of resonator modes. In the ultrastrong coupling regime the degeneracy is protected with respect to random fluctuations of the transition energies of the Josephson elements. 
 \end{abstract} 
 \maketitle 
Circuit quantum
electrodynamics (circuit-QED) is a very fascinating topic for
fundamental condensed matter physics, quantum optics and quantum
information. In superconducting circuit-QED systems, it has been
possible to implement on a chip the celebrated Jaynes-Cummings
model by strongly coupling a superconducting artificial atom to a bosonic
mode of a microwave transmission line resonator\cite{Wallraff,Yale_ladder} and to perform quantum logical operations with two
qubits\cite{DiCarlo}. So far, experimental manipulation of quantum states in
such circuit-QED systems has dealt with excited states. In these systems, the quantum 
ground state is non-degenerate and no information can be stored or processed by using only the vacuum of the circuit-QED system. In principle, by a judicious choice of their components,
superconducting quantum circuits can give rise to Hamiltonians, which cannot be achieved in atomic cavity-QED systems: in particular it may be possible to tailor the relative amplitude and the form of the interaction terms for the generation of interesting and controllable quantum vacuum properties.  

Here, we present a rigorous quantum field derivation showing that it is possible to obtain a vacuum degeneracy of a circuit-QED system by using a chain of Josephson junction atoms inductively coupled to a transmission line resonator.  A quantum critical coupling occurs in such a circuit-QED system thanks to both the type and ultrastrong size of the interaction obtainable with the inductive coupling scheme. In the case of a finite number $N$ of artificial atoms,  a degeneracy lifting occurs, with an energy splitting dramatically decreasing as $\exp(-g^2 \beta(N) )$, where $g$ is the dimensionless vacuum Rabi coupling  per atom (i.e., vacuum Rabi frequency divided by the transition frequency $\omega_F$)  and $\beta(N)  $  depends quadratically on $N$.  We present the asymptotic formula of the two degenerate vacua in the ultrastrong coupling limit .  Moreover, we show that the degeneracy is protected with respect to random site-dependent fluctuations of the Josephson transition energy. 
\label{Ham} 
\begin{center}
\begin{figure}[t]
\includegraphics[width=260pt]{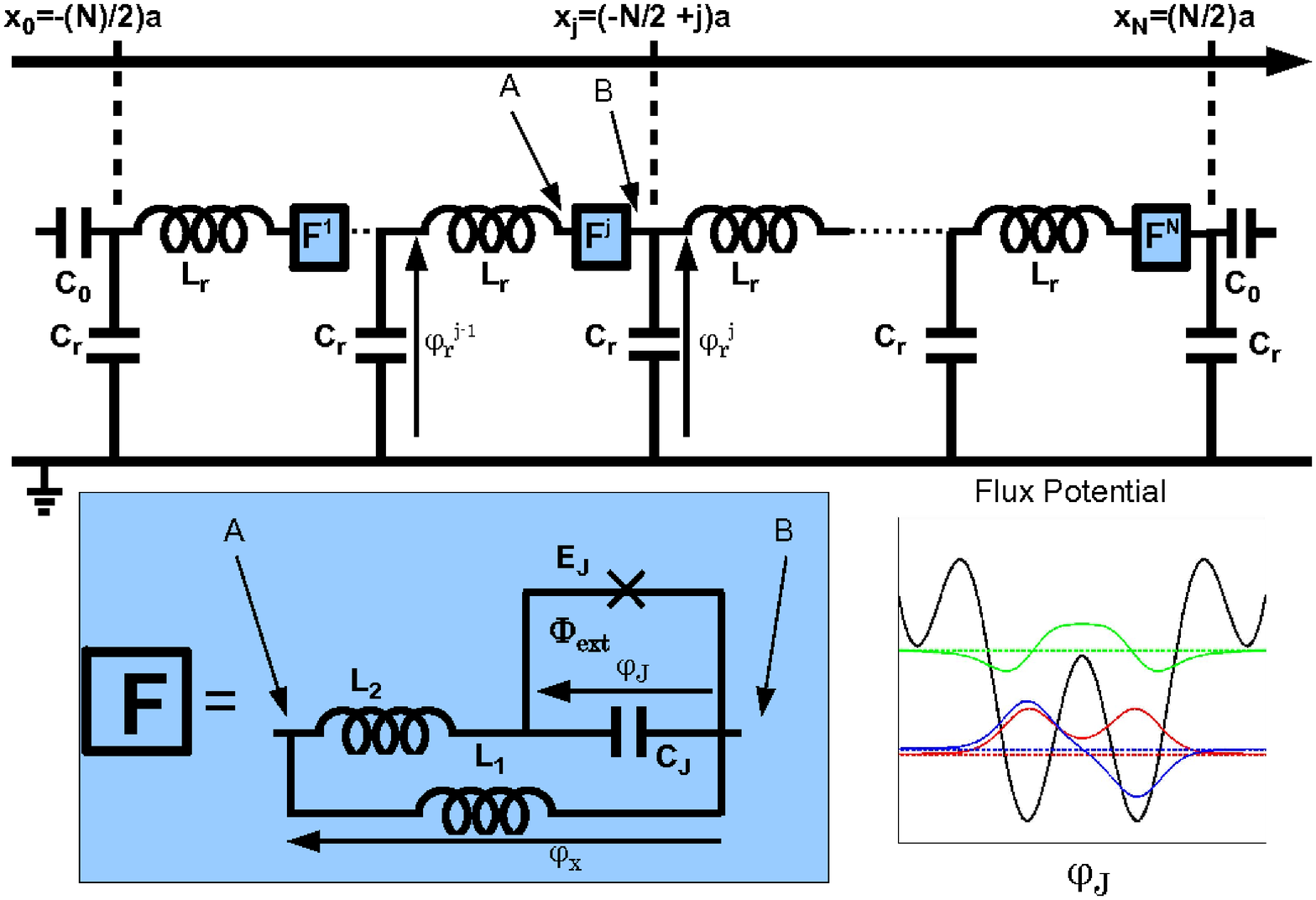}
\caption{\label{sketch}  A chain of $N$
 Josephson  atoms ("F" stands for
fluxonium\cite{fluxonium}) are inductively coupled to a transmission line resonator.  By tuning the external magnetic flux, the
flux-dependent potential for each fluxonium has a symmetric double
well structure with two states $\vert 0 \rangle$ and
$\vert 1 \rangle$ (with energy difference $\hbar \omega_F$) forming the two-level  system (parameters used for the inset: $E_J/E_{C_J} = 3$, $E_J/E_{L_J} = 20$).	}
\end{figure}
 \end{center}
A sketch of the proposed system is depicted in Fig.
\ref{sketch}, namely a chain of $N$ identical 
artificial two-level atoms in a transmission line
resonator. Each artificial atom ('fluxonium'  \cite{fluxonium}) is made of a Josephson junction coupled to inductances and an external magnetic flux. Here, we consider a scheme where each fluxonium is inductively coupled to the resonator. The fluxonium is known  be free from charge offsets\cite{fluxonium}; moreover, the inductive coupling can produce extremely large coupling even with a single artificial atom\cite{Devoret_strong}. In the case
of a chain, the Hamiltonian is $H=\sum_{j} H_j$ where each cell of
size $a$ is labeled by the index $j$ and is located at the
position $x_j$. One can effectively model the resonator as a
sequence of inductances $L_r=a l_r$ and capacitances $C_r=a c_r$
\cite{blais}, where $l_r$ ($c_r$) is the inductance (capacitance)
per unit length. The Hamiltonian of each cell reads:
\begin{eqnarray}
H_j=4E_{C_r} (\hat{N}_r^{j-1})^2\,+\,\frac{(\hat{\phi}_r^{j-1}-\hat{\phi}_r^{j}-\hat{\phi}^j_x)^2}{2L_r}+ \frac{(\hat{\phi}^j_x)^2}{2L_1}  \nonumber\\
+ \frac{(\hat{\phi}^j_x-\hat{\phi}^j_J)^2}{2L_2}\, +\,4E_{C_J}
(\hat{N}^j_J)^2-E_J \cos( {\frac{2e}{\hbar}} (\hat{\phi}^j_J +\Phi^j_{ext}))~,
\label{cell}
\end{eqnarray}
where $\hat{N}$ and $\hat{\phi}$ are the
number and flux operators for the resonator elements and Josephson
junctions ('$r$' stands for resonator; '$J$' for
Josephson junction). The charging
energies are $E_{C_r}=\frac{e^2}{2C_r}$ and
$E_{C_J}=\frac{e^2}{2C_J}$ . By applying Kirchoff's laws and by
taking $\Phi_{ext} = \pi \frac{\hbar}{2e}$, we find $H=H_{res}+H_{F}+H_{coupling}$, with:
\begin{eqnarray}
\label{H}
H_{res}=\sum_{j=1}^N 4E_{C_r} (\hat{N_r}^j)^2+E_{L_r}\frac{(\hat{\varphi}_r^j - \hat{\varphi}_r^{j-1})^2}{2}~, \nonumber\\
H_{F}=\sum_{j=1}^N 4E_{C_J} (\hat{N}^j_J)^2+E_{L_J}\frac{(\hat{\varphi}_{J}^j)^2}{2} + E_J \cos(\hat{\varphi}_{J}^j)~,\nonumber\\
H_{coupling}= \sum_{j=1}^N G(\hat{\varphi}_r^j - \hat{\varphi}_r^{j-1})\hat{\varphi}_J^j,
\end{eqnarray}
where we have introduced the dimensionless fluxes ${\varphi}_r^j =
\frac{2e}{\hbar} \hat{\phi}^j_r$,  $\varphi^j_J  =
\frac{2e}{\hbar}\hat{\phi}^j_J$ and the  inductance energy
constants are  $E_{L_r}=(\frac{\hbar}{2e})^2 \frac{ L_1 + L_2}{L_1
L_r\,+\,L_1 L_2\,+L_2 L_r}$ , $\,E_{L_J}=(\frac{\hbar}{2e})^2
(\frac{ L_1 + L_r}{L_1 L_r\,+\,L_1 L_2\,+L_2 L_r}) $ .
 The magnitude of the coupling constant is  $G=(\frac{\hbar}{2e})^2 \frac{L_1}{L_1 L_r\,+\,L_1 L_2\,+L_2 L_r} $.
 The  Hamiltonian $H_{res}$ describes the transmission line
resonator with a renormalized inductance per unit of length
$\tilde{l_r}= l_r \frac{L_1\,+\,L_2 + \frac{L_2 L_1}{a l_r}}{L_1 +
L_2}$, accounting for the additional inductances in each
fluxonium. By following the treatment in Ref.
\cite{blais}, the position-dependent flux field is
$
 \hat{\phi}(x)= i\sum_{k ~\geq 1} \frac{1}{\omega_k}\sqrt{\frac{\hbar
\omega_{k} }{2 c_r}}   f_k(x)
\,(\hat{a}_{k}\,-\hat{a}_{k}^{\dag})\,
$
where $a_{k}^{\dag}$ is the bosonic creation operator of a photon
mode with energy $\hbar \omega_k= \frac{k \pi a}{d} \sqrt{8E_{C_r}
E_{L_r}}$. The spatial profile of the $k$-th mode is $f_k(x)
=-\sqrt{2/d} \sin(\frac{k \pi x}{d})$ for $k$ odd, while $f_k(x)
=\sqrt{2/d} \cos(\frac{k \pi x}{d})$ for $k$ even, $d$  being  the
length of the one-dimensional resonator (in the following, we will consider $d = N a$). The site-dependent fluxes
are simply given by  $\hat{\phi}_r^j =
\hat{\phi}(x_j)$.
 
The Hamiltonian $H_{F}$ describes the sum of the energies of
the  fluxonium atoms. By properly tuning the external
 magnetic flux, it it possible to obtain a symmetric flux-dependent
potential energy, as shown in Fig. \ref{sketch}, with a
double well structure. Due to the strong anharmonicity of its energy
spectrum, the fluxonium can be approximated as a two-level system,
when $E_{J} \gg E_{L_J} $. We call the two first eigenstates of
the $j$-th fluxonium as $|0\rangle_j$ and $|1\rangle_j$ and we
introduce the raising operator $\hat{\sigma}_{+,j}= |1\rangle
\langle0|_j$ and $\hat{\sigma}_{-,j} =
\hat{\sigma}^{\dagger}_{+,j}=|0\rangle \langle 1|_j$. By using the
Pauli matrix notation, we have $\hat{\sigma}_{x,j} =
\hat{\sigma}^{\dagger}_{+,j} + \hat{\sigma}_{+,j}$ and
$\hat{\sigma}_{y,j} = i (\hat{\sigma}^{\dagger}_{+,j}
-\hat{\sigma}_{+,j})$ and $\hat{\sigma}_{z,j} = 2
\hat{\sigma}_{+,j}  \hat{\sigma}^{\dagger}_{+,j} - 1$. Leaving
aside a constant term, we then have $H_{F} = \sum_j \hbar \omega_F \frac{1}{2}
\hat{\sigma}_{z,j}$, where $\hbar \omega_F$
 is the energy difference between the two states $\vert 0 \rangle$ and $\vert 1 \rangle$. By considering only the
two-level subspace, the Josephson junction flux has the form
\begin{equation}
\hat{\varphi}^j_J \simeq
\langle 0 \vert \hat{\varphi}^j_J \vert 1 \rangle (\hat{\sigma}_{+,j}+\hat{\sigma}_{+,j}^{\dag}) = -\varphi_{01} \hat{\sigma}_{x,j},
\end{equation}
where $\varphi_{01} \simeq \pi$ for typical parameters ( see Fig. 1). 

As it will be clear in the following, it is convenient to
introduce excitation creation operators $\hat{b}^{\dagger}_{k}= \sqrt{\frac{2}{N}} \sum_{j=1}^{N} \Delta
f_{k}(x_{j})\hat{\sigma}_{+,j}$
 for $1\leq k \leq N-1$,
where $\Delta f_{k}(x_j)=\cos(\frac{k\pi(-\frac{N+1}{2}+j)}{d}a)$  for $k$ odd , and $\Delta f_{k}(x_j)=\sin(\frac{k\pi(-\frac{N+1}{2}+j)}{d}a)$  for $k$ even.
Note that the collective operator $\hat{b}^{\dagger}_{k}$ is a linear
superposition of the excitation operators in each fluxonium with
an amplitude depending on the spatial profile of
 the flux field of the resonator. In order to get a unitary transformation, it is
also necessary to introduce the operator
$\hat{b}^{\dagger}_N=\frac{1}{\sqrt{N}}\sum_j(-1)^j \hat{\sigma}_{+,j}$. In the
following, we will consider only the resonator modes $1\leq k \leq
N$, because, in the
conditions we are considering, the higher order (Bragg) modes are
energetically well off-resonant. Hence, we get the following effective
Hamiltonian:
\begin{equation} \label{Heff} {\mathcal
H}=\frac{\hbar}{2}\,\sum_{1\leq {k} \leq
N}\,\hat{\Phi}^{\dag}_{k}\,\eta\,\mathcal{M}_{k}\,\hat{\Phi}_{k}\,
\end{equation}
 where
 $\hat{\Phi}_{k}=(\,\hat{a}_{\,k},\,\hat{b}_{k},\,\,\hat{a}^{\dag}_{\,k},\,\hat{b}^{\dag}_{k})^{T}$
 with the Bogoliubov diagonal metric $\eta\,=\,diag[1,1,-1,-1]$, and
 the matrix:
\begin{equation}
\label{HB}
  \mathcal{M}_{k}=\left (
  \begin{array}{cccc} \omega_k & -i \Omega_k& 0 &-i \Omega_k\\ i \Omega_k &  \omega_{F}
  &  -i \Omega_k& 0 \\ 0& -i \Omega_k& -\omega_k &  -i \Omega_k \\  -i\Omega_k& 0 & i\Omega_k& - \omega_{F} \end{array}
  \right).
\end{equation}
The coupling between the annihilation operators 
$\hat{a}_{\,k}$, $\hat{b}_{k}$ and the creation operators are due to the antiresonant (non-rotating wave) terms present in the coupling Hamiltonian $H_{coupling}$.
The collective vacuum Rabi frequency reads  for $1\leq k \leq N-1$
\begin{equation}
\label{omegaq}
  \hbar \Omega_k=G \frac{4e}{\hbar} \varphi_{01} \sin (\frac{k \pi a}{2d})\frac{1}{\omega_k}\sqrt{\frac{\hbar
\omega_{k} N }{2 d c_r}}.
  \end{equation}
( and for $k=N$, $\hbar
\Omega_N= G \frac{4e}{\hbar} \frac{\varphi_{01}}{\omega_N}\sqrt{\frac{\hbar
\omega_{N} N }{d c_r}}$).\\
 Notice that each $k$-mode of the resonator is coupled only to the collective matter mode
with the same spatial symmetry and ${\mathcal H} = \sum_k
{\mathcal H}_k$. Hence, the eigenstates are products
of the eigenstates corresponding to the $k$-subspaces.  The
effective Hamiltonian in Eq. (\ref{Heff}) has been obtained by
assuming that the operators $\hat{b}_{k}^{\dag}$ are
 bosonic, i.e. $[\hat{b}_{k}, \hat{b}_{k} ^{\dag}]
\simeq 1 $, an approximation working in the limit $N \gg 1$. 
The excitation spectrum of the collective 
bosonic modes depends on the eigenvalues of the matrix
$\mathcal{M}_{k}$. A crucial property is given by the determinant $Det(\mathcal{M}_{k})=\omega_k\omega_F(\omega_k\omega_F-4\Omega_k^2)$,  which vanishes when the vacuum Rabi frequency equals the critical value
$
\Omega^c_k = \frac{ \sqrt{\omega_k \omega_{F}}}{2}
$,
implying that two of the 4 eigenvalues of $\mathcal{M}_{k}$ are
exactly zero. For $\Omega_k > \Omega_c^k$, two of the 4 eigenvalues of the
matrix $\mathcal{M}_{k}$ becomes imaginary, manifesting an instability of
the normal, non-degenerate, quantum vacuum phase. \\This is reminiscent of quantum phase transitions\cite{QPT} with
Dicke-like Hamiltonians\cite{Brandes}, where at the quantum
critical point there is a gapless bosonic excitation. 
Note that Dicke-like Hamiltonians are usually
obtained by dropping the so-called $\mathbf{A}^2$-term, that is the term
associated to the squared electromagnetic vector potential term \cite{Nori}, \cite{Dimer}. 
However, because of the magnitude of the $\mathbf{A}^2$-term, a system with (ultra)strong light-matter coupling does not necessarily have a quantum critical point\cite{note_D}. For example, in the celebrated Hopfield model\cite{Hopfield} for dielectrics a quantum critical point does not exist even if the coupling can be ultrastrong \cite{Ciuti_vacuum,Ciuti_PRA}.
In our case, the quantum field Hamiltonian in Eq. (\ref{H}) is the complete Hamiltonian for the proposed superconducting system and no term has been omitted. In particular, the analogous of the $\mathbf{A}^2$ term in the present system is given by extra terms proportional to $(\hat{\varphi}_r^j - \hat{\varphi}_r^{j-1})^2$ obtained from Eq. (1) after substitution of the expression for the flux $\hat{\phi}_x^j$ obtained by Kirchoff's laws\cite{Kirch}. Here, these terms are fully included and contribute to the expression for the resonator renormalized inducting energy $E_{L_r}$, which does depend on $L_1$ and $L_2$. 
\begin{center}
\begin{figure}[t!]
\includegraphics[width=270pt]{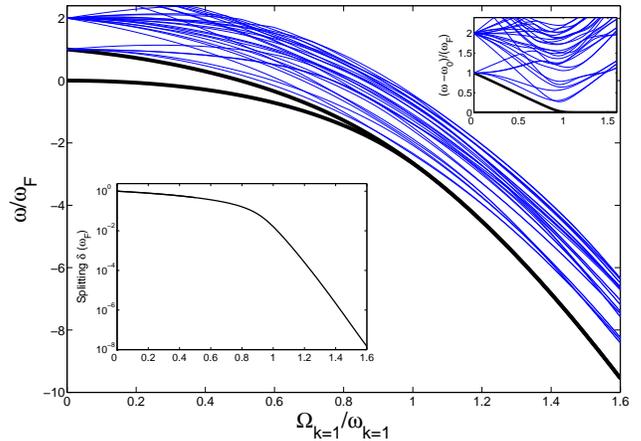}
\caption{\label{numerical} First 30 energy eigenvalues versus dimensionless
vacuum Rabi coupling for $N=5$ fluxonium atoms, $N_m=3$ resonator modes ($\omega_F = \omega_{k=1}$). Upper inset: the difference between the energy eigenvalues and the ground state energy is plotted.
Lower inset: normalized energy difference (log scale) between the first 2 quasi-degenerate levels versus the dimensionless coupling.
In the ultrastrong coupling limit, the two ground states are excellently
approximated by the analytical formula in Eq. (\ref{exact}) ($98 \%$ overlap for the largest coupling here considered).
}
\end{figure}
 \end{center}
For the case of finite number of fluxonium atoms $N$ and finite number of modes $N_m$, we have performed numerical diagonalizations of the circuit-QED Hamiltonian. As shown in Fig. 2, for increasing coupling the energy of the first excited state converges towards the energy of the ground state, hence a twice degenerate vacuum is obtained in the ultrastrong coupling limit. As shown in the inset, for a given value of $N$, the energy splitting exponentially decreases with increasing vacuum Rabi coupling.
The finite-size scaling properties are shown in Fig. 3(a), where the energy splitting is plotted as a function of $g^2 $, where $g=\frac{\Omega_{k=1}}{\sqrt{N} \omega_{k=1}}$ is the dimensionless vacum Rabi coupling frequency per fluxonium. Our numerical  results show that the energy splitting scales $\exp(-g^2 \beta(N) )$, where $\beta(N) \approx 2 N^2$ (see inset of Fig. 3a). Hence, a perfect degeneracy is obtained either in the thermodynamical limit ($N \to + \infty$) or for $g \gg 1$. As shown later, it is possible to have $g \gg 1$ in realistic superconducting systems, hence a negligible splitting can be achieved with a relatively small $N$.
In the ultrastrong coupling limit ($\frac{\Omega_{k=1}}{\omega_F} \rightarrow
\infty$), we have derived\cite{additional} an analytical expression for the two degenerate ground states
by taking into account an arbitrary number $N_m$ of modes  for the  resonator.  It is convenient to introduce the $x$-polarized
states (eigenstates of $\hat{\sigma}_{x,j}$), namely $ |+\rangle_j
= \frac{1}{\sqrt{2}}(|1\rangle_j + |0\rangle_j)$ and $|-\rangle_j
= \frac{1}{\sqrt{2}}(|1\rangle_j - |0\rangle_j)$. We have found 
that in the ultrastrong coupling limit (where $H_F$ is dominated by  $H_{res} + H_{coupling})$
the asymptotic expression for the two degenerate vacua  $\vert G_{+} \rangle$ and $\vert G_{-} \rangle$ is:
\begin{eqnarray}
\label{exact}
|G_{\pm}\rangle = C_{G}\Pi_j|\pm\rangle_j\otimes \Pi_{k_o} e^{ \pm (\frac{g\sqrt{2}~i^{k_o}}{k_o^{1.5} \sin(\frac{\pi }{2N})}   a_{k_o}^{\dag})} |0\rangle_{k_o} \otimes \Pi_{k_e} |0\rangle_{k_e}\nonumber\\
\end{eqnarray}
with $C_{G}$ a normalisation constant, $k_o$ ($k_e$)
standing for the odd (even) $k$ values for the resonator modes. Eq. (\ref{exact}) shows that the two degenerate ground states are
the product of a 'ferromagnetic' state for the chain of artificial atoms
times coherent states for the different resonator modes.
Importantly, the two orthogonal ground states have opposite
polarization of the pseudospins and opposite phases for the coherent
states. Due to the mode spatial symmetry, in $|G_{\pm}\rangle$ the even $k_e$ resonator modes are empty.   The analytical expression for the two vacua excellently agrees with the numerical results.  

It is interesting to see how the degeneracy is affected by the presence of an additional Hamiltonian term $H_{pert} = \sum_{j} \frac{1}{2}\hbar \Delta_j  \hat{\sigma}_{z,j}$, describing a site-dependent random fluctuation of the fluxonium energies.  Interestingly, we have found numerically (see Fig. 3b) and analytically\cite{additional} that the average splitting $<\delta>$ and its standard deviation $\sigma= \sqrt{<\delta^2>-<\delta>^2}$ have the same exponential dependence as the splitting in
the ideal case of identical Josephson elements. By having $N$ and/or $g$ large enough, the effect of a disorder of given amplitude can be made arbitrary small.  This occurs because  $\langle G_{\pm} \vert H_{pert}^m \vert \hat
G_{\pm}\rangle = \langle  G_{\pm} \vert H_{pert}^m \vert
G_{\mp}\rangle = 0$ with $m \leq N-1$, i.e.,  such a perturbation is zero up to the $N$-th order perturbation theory, leading to a protected degeneracy\cite{Doucot}.
\begin{center}
\begin{figure}[t]
\includegraphics[width=270pt]{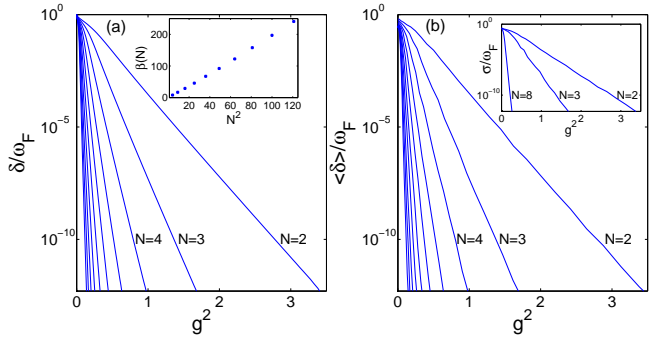}
\caption{\label{numerical} Normalized energy splitting (log scale) between the two quasi-degenerate vacua versus $g^2$ for different values of $N$ (2,3,4,...11), where $g$ is the dimensionless vacuum Rabi coupling frequency per artificial atom. (a) Results for the case of identical fluxonium atoms with same transition frequency $\omega_F$. $ \delta/ \omega_F$ decreases as $\exp(-\beta(N)g^2)$. Inset: $\beta(N)$ versus $N^2$. (b) Averaged splitting $<\delta>$ in presence of a random distribution: $\omega_{j,F} = \omega_F + \Delta_j = \omega_F (1 + 0.5 \xi_j)$, where
$j$ labels the site and $\xi_j$ is a gaussian variable with variance equal to 1. The results have been averaged over $100$ disorder configurations. The standard deviation $\sigma=\sqrt{<\delta^2>-<\delta>^2}$ has the same exponential dependence (see inset).
}\end{figure}
 \end{center}

The present system can indeed have 
a very large controllable coupling. For $\omega_F=\omega_{k=1}$,  we have
\begin{eqnarray}
\frac{\Omega_{k=1}}{\omega_{k=1}}=  g \sqrt{N} =  \sqrt{\frac{Z_{vac}}{2 Z_r \alpha}} \mu \nu \chi\sqrt{ N
} \sim 5.7
 \chi \sqrt{ N }~,
\end{eqnarray}
where $\nu=\frac{1}{4\pi}\varphi_{01} \sim \frac{1}{4}$ for
$\frac{E_J}{E_{L_J}}\gg 1$, $\mu=\frac{\sin (\frac{\pi
a}{2d})}{\frac{\pi a}{2d}}$. For $\frac{a}{d}\rightarrow 0^+$, we have $\mu \sim 1$. Moreover,
$\frac{Z_{vac}}{2  \alpha}=\frac{h}{e^2}=R_k \sim 25.8 k \Omega$ is the impedance quantum, while $Z_r=\sqrt{\frac{L_r}{C_r}}=50\Omega$ is the standard transmission line impedance .
Finally, the branching ratio $\chi=(\frac{L_r}{L_1L_r+L_1L_2+L_2L_r})^{\frac{1}{4}}\frac{L_1}{(L_1+L_2)^{\frac{3}{4}}}$ is the control parameter to tune  $\frac{\Omega_{k=1}}{\omega_{k=1}}$.  $\chi \simeq 0$ for $L_1 \ll L_2$ and $\chi \simeq 1$ when
$\frac{L_1}{L_2} \gg 1$. 

Note that the bosonic mode population  in the ground state cannot give rise to any extracavity microwave radiation unless a non-adiabatic modulation of the Hamiltonian
is applied\cite{Ciuti_vacuum,Ciuti_PRA,Simone_PRL}. The quantum vacuum radiation across the quantum phase transition is an interesting problem to explore in the future. In the opposite limit of adiabatic changes of the Hamiltonian, thanks to the degeneracy, it may be possible to create non-abelian Berry phases and control quantum superpositions in the ground state subspace (a sort of vacuum qubit).  The system studied here appears promising for the observation of quantum phase transitions and the manipulation of quantum vacua in circuit-QED.

We are grateful to M.H. Devoret for many discussions and for suggesting us to consider the fluxonium atom. We thank B. Dou\c cot, I. Carusotto, S. De Liberato for discussions and M. Bamba for numerical advice.

\begin{widetext}

\section{Supplementary online material for the paper :
\textit{Vacuum degeneracy of a circuit-QED system in the ultrastrong coupling regime }}

In Section A of this Supplementary Material, we give a detailed proof of our analytical expression for the two degenerate vacua in the ultrastrong coupling regime. In Section B, we show the analytical calculation of the degeneracy splitting in the ultrastrong coupling limit. We show also useful details about the degeneracy with respect to site-dependent fluctuations of the Josephson atom transition frequencies (energy disorder).
 
\section{A. Ultra-strong coupling limit: analytical approach}
Let us consider the ultrastrong coupling limit ($\frac{\Omega_{k=1}}{\omega_F} \rightarrow + \infty$) , by taking into account an arbitrary number of modes $N_m$  for the transmission line resonator and a finite number $N$ of fluxoniums qubits.  The following treatment is general and can be done for any kind of mode spatial profile $\Delta f_{k}(x)$. In the ultra-strong coupling regime, the bare Hamiltonian  $H_{F}$ of the pseudospins (the Josephson artificial atoms) can be treated as a perturbation of the Hamiltonian $H_{res} + H_{coupling}$, where $H_{res}$ is the Hamiltonian of the bare transmission line and $H_{coupling}$ is the interaction part between the resonator and the artificial atoms.
As written in the letter,  it is convenient to consider the  basis $(|-\rangle, |+\rangle)_j$, the two eigenvectors of the operator $\hat{\sigma}_{x,j}$. The Hamiltonian can be rewritten in the form:
\begin{equation}
H_{res} + H_{coupling} =\sum_{k=1..N_m} \hbar \omega_{k}  a_{k}^{\dag} a_{k}+ \sum_{k=1..N_m} \sum_{j = 1} ^{N} i  \hbar  \Omega_k  \sqrt{ \frac{ 2}{N}} \Delta f_{k}(x_j) (a_{k}\,-a_{k}^{\dag})   ( (|+\rangle \langle +|_j -|-\rangle \langle -|_j  )
\end{equation}
Let us introduce the subspace $\mathcal{F}_{S_{\zeta}}$ generated by the states $ \Pi_j|\zeta_j\rangle \otimes \vert \Psi_{res}   \rangle \,$ where   the $\zeta_j \in \{-,+ \}$ denote a given pseudospin configuration sequence $S_{\zeta}$ for the chain, while $ \vert \Psi_{res}  \rangle $ describes a generic state of the resonator bosonic quantum field. Since there are $N$ two-level systems, then we have $2^N$  subspaces $\mathcal{F}_{\{|\zeta_j\rangle\}_{j=1..N}}$, each for every pseudospin configuration. 

It is clear that  by applying $H_{res} + H_{coupling}$ on a state $\vert \psi\rangle \in \mathcal{F}_{\{|\zeta_j\rangle\}_{j=1..N}}$, the result is a state belonging to the same subspace (i.e., $H_{res} + H_{coupling}$ conserves a given pseudospin configuration).  Now, if we consider a given pseudospin configuration sequency $S_{\zeta}$ for the chain,  
we can define $\psi^{S_{\zeta}}_k=\sum_j \sqrt{ \frac{ 2}{N}} \Delta f_{k}(x_j) \mu_{S_{\zeta}}(j)$
 where $\mu_{S_{\zeta}}(j)=1\,$if $|S_{\zeta_j}\rangle = |+\rangle_j \,$ and where $\mu_{S_{\zeta}}(j)=-1\,$if $|S_{\zeta_j}\rangle = |-\rangle_j \,$, we have that the Hamiltonian on the $(\mathcal{F}_{S_{\zeta}})$ can be written as:
\begin{equation}
H_{res}^{S_{\zeta}}+H_{coupling}^{S_{\zeta}}\\
=\sum_{k=1..N_m}(\hbar \omega_{k} a_{k}^{\dag} a_{k}+i \hbar \Omega_{k}   (a_{k}\,-a_{k}^{\dag}) \psi^{S_{\zeta}}_k )\\
=\sum_{k=1..N_m}(\hbar \omega_{k} (a_{k}^{\dag}+i\frac{\Omega_{k}}{\omega_{k}}\psi^{S_{\zeta}}_k )(a_{k}-i\frac{\Omega_{k}}{\omega_{k} }\psi^{S_{\zeta}}_k ) - \hbar \frac{\Omega_{k}^2}{\omega_{k}}(\psi^{S_{\zeta}}_k )^2).
\end{equation}
So if we introduce the shifted boson operator $\tilde{a}_k^{S_{\zeta}}=a_{k}-i\frac{\Omega_{k}}{\omega_{k} }\psi^{S_{\zeta}}_k$, we have the same bosonic commutation relations between ${a}_k^{S_{\zeta}}$ and $ (\tilde{a}_k^{S_{\zeta}})^{\dag}$ and
the Hamiltonian terms read \begin{eqnarray}
H^{S_{\zeta}}=\sum_{k=1..N_m}\hbar \omega_{k} (\tilde{a}_k^{S_{\zeta}})^{\dag}\tilde{a}_k^{S_{\zeta}} -\sum_{k=1..Nm} \hbar \frac{\Omega_{k}^2}{\omega_{k}}(\psi^{S_{\zeta}}_k )^2
\end{eqnarray}
So, on this subspace, the fundamental state $|G_{S_{\zeta}}\rangle$ has the energy $E_{G_{S_{\zeta}}} = -\sum_{k=1..N_m} \hbar \frac{\Omega_{k}^2}{\omega_{k}}(\psi^{S_{\zeta}}_k )^2$ .
Moreover, we have: $$\tilde{a}_k^{S_{\zeta}}|G_{S_{\zeta}}\rangle=0\,\,\,\forall k \leq N_m$$
This relation implies that:
\begin{equation}
|G_{S_{\zeta}}\rangle=\Pi_j|S_{\zeta_j}\rangle\otimes \Pi_k e^{-\frac{(\frac{\Omega_{k}}{\omega_{k} }\psi^{S_{\zeta}}_k)^2}{2}}e^{ (i\frac{\Omega_{k}}{\omega_{k} }\psi^{S_{\zeta}}_k a_k^{\dag})} |0\rangle_k \\
\end{equation}
In order to find the ground state for the complete Hilbert space, we have to determine the pseudospin configuration minimizing $-\sum_{k=1..N_m} \hbar \frac{\Omega_{k}^2}{\omega_{k}}(\psi^{S_{\zeta}}_k )^2=- 2 g^2 (\hbar\omega_{k=1})\sum_{j,j'}  \mu_{S_{\zeta}}(j) Q(j,j')\mu_{S_{\zeta}}(j')$ where $g=\frac{\Omega_{k=1}}{\sqrt{N} \omega_{k=1}}$ and where we have called Q  the quadratic form 
$$Q(j,j')= \sum_{k=1..N_m} \Delta f_{k}(x_j)[ (\frac{\Omega_k}{\Omega_{k=1}})^2\frac{1}{k} ]\Delta f_{k}(x_j'). $$
For a given number of modes $N_m$ and a given spatial profile $\Delta f_{k}(x)$ and position $(x_j)_{j=1..N}$, we derive the form and we find beyond the $2^N$ configurations $S_{\zeta}$, the one which minimizes $-\sum_{j,j'}  \mu_{S_{\zeta}}(j) Q(j,j')\mu_{S_{\zeta}}(j')$.
In fact the double degeneracy of the spectrum appears also here because at any configuration $S_{\zeta}$ corresponds, an opposite one $S_{\zeta'}$
(for which $\mu_{S_{\zeta'}}(j) =-\mu_{S_{\zeta}}(j)\,\,\,\,\,\,\forall j$) with same energy.
With our particular profile , the configurations of minimal energy are the two ferro-magnetic ones:
$\mu_{S_{\zeta}}(j)=+1\,\,\,\forall j$ and $\mu_{S_{\zeta'}}(j)=-1\,\,\,\forall j$.
So, to conclude, the two fundamental states we derived are :

\begin{eqnarray}
|G_{\pm}\rangle=\Pi_j|{\pm}\rangle\otimes \Pi_k e^{-\frac{(\frac{\Omega_{k}}{\omega_{k} }\psi^{S_{{\pm}}}_k)^2}{2}}e^{ (i\frac{\Omega_{k}}{\omega_{k} }\psi^{S_{{\pm}}}_k a_k^{\dag})} |0\rangle_k =C_{G}\Pi_j|\pm\rangle_j\otimes \Pi_{k_o} e^{ \pm (\frac{g\sqrt{2}~i^{k_o}}{k_o^{1.5} \sin(\frac{\pi }{2N})}   a_{k_o}^{\dag})} |0\rangle_{k_o} \otimes \Pi_{k_e} |0\rangle_{k_e}
\end{eqnarray} 
where $C_{G}= \Pi_{k_o} e^{ -\frac{(\frac{g\sqrt{2}}{k_o^{1.5} \sin(\frac{\pi }{2N})})^2}{2}}$  and with $k_o$ ($k_e$)
standing for the odd $k$ values for the resonator modes.

\section{B: Degeneracy splitting and protection with respect to site-dependent energy disorder in the ultrastrong coupling limit}

Now let us consider the effect of  the bare Hamiltonian for the artificial atoms, namely $H_{F}= \hbar \omega_F \sum_{j=1}^N  \frac{1}{2} \hat{\sigma}_{z,j} = \hbar \omega_F \sum_{j=1}^N   \frac{1}{2}(|+\rangle \langle -|_j +|-\rangle \langle +|_j) $, 
where $|\pm\rangle_j$ are the eigenstates of $\hat{\sigma}_{x,j}$. In the ultrastrong coupling limit, $H_F$ acts as a perturbation of $H_{res} + H_{coupling}$. In the finite-size case, this produces a degeneracy splitting for the two vacua $|G_{+}\rangle$ and $|G_{-}\rangle$. In presence of  $N$ artificial atoms, the effect of $H_F$ is zero up to the $N$-th order in perturbation theory. As shown by the numerical results reported in our letter,  the splitting decreases as $\sim e^{-\beta(N) g^2}$ with $\beta(N) \approx 2 N^2$ (see inset of Fig. 3a). Here we show in detail for the case $N=2$ that this exponential dependence can be found analytically. Moreover, the same result occurs in presence of a random site-dependent fluctuation of the artificial atom transitions energies (energy disorder). 
We finally discuss the N$\geq 2$ case.

\subsection{Degeneracy splitting for N=2 fluxoniums in the ultrastrong coupling limit}
Let us consider the term $ \hbar \omega_F \frac{1}{2} (|+\rangle \langle -|_1 +|-\rangle \langle +|_1)$, that is the bare Hamiltonian term associated to the first artificial atom.
Taken alone, such a term does not lift he degeneracy since it does not couple directly $|G_{+}\rangle$ and $|G_{-}\rangle$ at any order. However, it produces a mixing of the vacuum states $|G_{+}\rangle$ and $|G_{-}\rangle$ with the excited states of $H_{res} + H_{coupling}$. Accordingly, we can introduce the following states
\begin{eqnarray}
|\tilde{G}_{+}\rangle  \simeq|G_{+} \rangle + \frac{1}{2} \hbar \omega_F \sum_{\mathbf{n}} \frac{\langle \mathbf{n}, -+ |\sigma^z_1\vert G_{+}\rangle }{E_{G_{+}}-E_{\bf{n},-+ }}  \vert \mathbf{n},-+ \rangle \\
  |\tilde{G}_{-}\rangle  \simeq |G_{-} \rangle +  \frac{1}{2} \hbar \omega_F \sum_{\mathbf{n}} 
\frac{\langle \mathbf{n},+- \vert \sigma^z_1\vert G_{-} \rangle }{E_{G_{-}}-E_{n,+- }} 
 | \bf{n},+- \rangle,
\end{eqnarray}
where $\mathbf{n} = (n_1,n_2,....,n_{N_m})$ and the states $ | \mathbf{n},-+ \rangle=\frac{1}{\sqrt{n_1 ! ...n_{N_m}!}} ((\tilde{a}_{k=1}^{{-+}})^{\dag})^{n_1}...((\tilde{a}_{k=N_m}^{{-+}})^{\dag})^{n_{N_m}} |G_{{-+}}\rangle$ stand for the excited  eigenstates of $H_{res} + H_{coupling}$ with pseudospin configuration $\{-+\}$. Their corresponding eigenenergies are $E_{\mathbf{n},-+} = \sum _{k=1, ..,N_m} n_k \hbar \omega_{k} + E_{G_{-+}} $
. They are obtained by applying the shifted photonic creation operators $(\tilde{a}_k^{{-+}})^{\dag} =a_k^{\dag}+i\frac{\Omega_{k}}{\omega_{k} }\psi^{-+}_k$ for the mode numbers $k=1$ to $k=N_m$.
Anagolous definition holds for $ | \mathbf{n},+- \rangle $, corresponding to the pseudospin configuration $\{+-\}$ .\\ 
Now, we can take care of  $ \hbar \omega_F \frac{1}{2}\sigma^z_2=\hbar \omega_F \frac{1}{2} \sigma_(|+\rangle \langle -|_2 +|-\rangle \langle +|_2) $, that is the contribution of the second pseudospin. The energy splitting between the vacua $|G_{+}\rangle$ and $|G_{-}\rangle$
at the $N=2$ order perturbation theory is
\begin{eqnarray}
\delta =   \omega_F\langle \tilde{G}_{-}| \sigma^z_2 |\tilde{G}_{+}\rangle
 =\frac{ \omega_F}{2}\left (\sum_{\mathbf{n}} 
   \frac{ \hbar \omega_F \langle G_{+}|\sigma^z_2| \mathbf{n},+- \rangle \langle \mathbf{n},+- |\sigma^z_1|G_{-} \rangle}{E_{G_{+}}-E_{\mathbf{n} ,+- }}  + \sum_{\mathbf{n}} 
   \frac{\hbar \omega_F \langle G_{-}|\sigma^z_2| \mathbf{n}, -+ \rangle \langle \mathbf{n},-+ |\sigma^z_1|G_{+} \rangle}{E_{G_{-}}-E_{\mathbf{n},-+ }} \right ) \nonumber \\
   =  \omega_F \sum_{\mathbf{n}} 
   \frac{\hbar \omega_F \langle G_{+}|\sigma^z_2| \mathbf{n},+- \rangle \langle \mathbf{n},+- |\sigma^z_1|G_{-} \rangle}{E_{G_{-}}-E_{\mathbf{n},+-}}. 
\end{eqnarray}
We can now consider the effect of only one photonic mode( $N_m=1$). In that case, the result is simplified because $\psi^{-+}_{k=1}= -\Delta f_{k=1} (x_1) +\Delta f_{k=1} (x_2) = 0$, hence the excited states $ |n,+- \rangle $ are simply the unshifted Fock states  $\vert n \rangle $. Then, the energy splitting reads:
\begin{eqnarray}
\delta \simeq   \frac{\omega_F^2}{\omega_{k=1}} C_G^2\sum_{n} \frac{\langle 0 | e^{ -2gi  a_{k=1})} | n \rangle \langle n | e^{ - 2gi  a_{k=1}^{\dag})} |0\rangle}{4g^2+n}  \nonumber \\
 =   \frac{\omega_F^2}{\omega_{k=1}} e^{-4g^2}\sum_{n} \frac{(-4g^2)^n}{(4g^2+n){n!}}
 \simeq  \frac{\omega_F^2}{2 \omega_{k=1}} \sqrt{\frac{\pi}{2g^2}} e^{-8g^2}.
 \end{eqnarray}
In the last derivation we used the identity $\psi^{++}_{k=1}=\Delta f_{k=1} (x_1) +\Delta f_{k=1} (x_2) = \sqrt{2}=-\psi^{--}_{k=1}$ and $E_{G_{+}}=E_{G_{-}}=-4\hbar \omega_{k=1} g^2$.
Note that we have verified that this analytical expression is an excellent approximation of the exact numerical results (for the parameters used in the figures of our manuscript the analytical approximation differs less than 10$\%$ from the numerical results).

\subsection{Degeneracy splitting for $N=2$ in presence of site-dependent energy disorder in the ultrastrong coupling limit}

Now, we wish to consider a configuration where the artificial atoms have not the same energy.
In Fig. 3b of our manuscript, we have reported numerical results showing that a disorder-induced degeneracy splitting dramatically decreases with the coupling per fluxonium $g$ and size $N$. In particular, we have shown that there is the same exponential dependence as in the absence of disorder. 
This can be proved analytically for the $N=2$ case.
We consider 
\begin{equation}
H_F = \sum_{j=1,2} \hbar \omega_{F,j} \frac{1}{2} \hat{\sigma}_{z,j}
\end{equation}
Following the same steps as in the previous derivation we find:
\begin{equation}
   \delta \simeq  \omega_{F,2} \sum_{\mathbf{n}} 
   \frac{ \hbar \omega_{F,1} \langle G_{+}|\sigma^z_2| \mathbf{n},+- \rangle \langle \mathbf{n},+- |\sigma^z_1|G_{-} \rangle}{E_{G_{-}}-E_{\mathbf{n},+-}}  \simeq \frac{\omega_{F,2}}{2} \frac{\omega_{F,1}}{\omega_{k=1}} \sqrt{\frac{\pi}{2g^2}} e^{-8g^2}.
\end{equation}
If we now consider different disorder realizations and we average over the configurations, we get an average splitting:
\begin{equation}
<\delta> \simeq  \frac{<{\omega_{F,1} \omega_{F,2}}>}{2 \omega_{k=1}}  \sqrt{\frac{\pi}{2g^2}} e^{-8g^2}.
\end{equation}
Taking  $\omega_{F,i} = \omega_F + \Delta_i$ with $\Delta_i$ a random variable with zero average ($<\Delta_j> =0$), variance $<\Delta_i^2> = \Delta^2$ and such that $<\Delta_i \Delta_{j \neq i} >=0$ \,\, $\forall i, j $ , then the averaged degeneracy splitting $<\delta>$ and the standard deviation $\sigma$ read:
\begin{eqnarray}
<\delta> \simeq   \frac{\omega_F^2}{2 \omega_{k=1}} \sqrt{\frac{\pi}{2g^2}} e^{-8g^2}, \\
\sigma=\sqrt{<\delta^2>-<\delta>^2} \simeq \sqrt{2 + \left (\frac{\Delta}{\omega_F} \right)^2 } \ \frac{\Delta}{\omega_F}  <\delta>. \end{eqnarray}
Hence, the averaged splitting is equal to the splitting without disorder for the case of identical Josephson elements. The standard deviation of the splitting
depends on the normalized disorder amplitude $\Delta/\omega_F$, but it has the same exponential dependence
as $<\delta>$.  Hence, in the ultrastrong coupling limit ($g \gg 1$), the effect of disorder can be made arbitrarily small.

Note that a similar protection occurs with respect to local noise sources proportional to $\hat{\sigma}_{y,j}$.

 \subsection{Degeneracy splitting for $N\geq2$ fluxoniums}
We now consider the general case $N \ge 2$ and Josephson artificial atoms with site-dependent energy, i.e.  $H_F = \sum_{j=1..N} \hbar \omega_{F,j} \frac{1}{2} \hat{\sigma}_{z,j}$. In  this general case, $H_F$ couples the two degenerate vacua only at the $N$-th order in perturbation theory. For finite values of $N$ and $g$, the energy splitting is given by the  following expression:
\begin{eqnarray}
 \hbar \delta \simeq 2\left [ \Pi_{j=1..N} (\frac{\omega_{F,j}}{2 \omega_{k=1}})  \right ] | \sum_{ \sigma \in \mathfrak{S_{\mathit{N}}}}  \sum_{\mathbf{n_1}, \mathbf{n_2}, ...\mathbf{n_{N-1}}}  \\
   \frac{ \langle G_{+}|\sigma^z_{\sigma(1)}| \mathbf{n_1}, _{\sigma(1)(S_+)} \rangle \langle \mathbf{n_1},_{\sigma(1)(S_+) }|\sigma^z_{\sigma(2)}|  \mathbf{n_2}, _{\sigma(2)(\sigma(1)(S_+) )}\rangle.... \langle \mathbf{n_{N-1}},_{\sigma(N-1)(...\sigma(1)(S_+))} |\sigma^z_{\sigma(N)}|  G_-\rangle}{(E_{G_{+}}-E_{\mathbf{n_1}, _{\sigma(1)(S_+)}})(E_{\mathbf{n_1}, _{\sigma(1)(S_+)}}-E_{\mathbf{n_2}, _{\sigma(2)(\sigma(1)(S_+) )}})...(E_{\mathbf{n_{N-2}}, _{\sigma (N-2)(..\sigma(1)(S_+))}}-E_{\mathbf{n_{N-1}}, _{\sigma (N-1)(...\sigma(1)(S_+) )}})} |, 
   \end{eqnarray}
where  $\mathfrak{S_{\mathit{N}}}$ is the set of permutations of $\{1..N\}$, $S_+$ is the $\{++...+\}$ configuration, and where $\sigma(m)(...\sigma(1)(S_+))$ stands for the pseudo-spin configuration in which the $\sigma(1)^{th}$, $\sigma(2)^{th}$ ... $\sigma(m)^{th}$ pseudo-spins have switched from + to -. The last expression contains all the excited states of every pseudo-spin configurations, with their energies at the denominator. In fact, in the ultrastrong coupling limit, the denominator will give a polynomial contribution to the splitting  proportional to $(\hbar \omega_{k=1})^{N-1}$. Hence, we get
\begin{eqnarray}
 \delta \sim 2  \omega_{k=1}\left [ \Pi_{j=1..N} (\frac{\omega_{F,j}}{2 \omega_{k=1}})  \right ]  \sum_{ \sigma \in \mathfrak{S_{\mathit{N}}}}  \sum_{\mathbf{n_1}, \mathbf{n_2}, ...\mathbf{n_{N-1}}} \\ \nonumber
   \langle G_{+}|\sigma^z_{\sigma(1)}| \mathbf{n_1}, _{\sigma(1)(S_+)} \rangle \langle \mathbf{n_1},_{\sigma(1)(S_+) }|\sigma^z_{\sigma(2)}|  \mathbf{n_2}, _{\sigma(2)(\sigma(1)(S_+) )}\rangle.... \langle \mathbf{n_{N-1}},_{\sigma(N-1)(...\sigma(1)(S_+))} |\sigma^z_{\sigma(N)}|  G_-\rangle\\ \nonumber= 2 \omega_{k=1} N!  \left [ \Pi_{j=1..N} (\frac{\omega_{F,j}}{2 \omega_{k=1}})  \right ] \langle G_{+}|\Pi_{j=1..N} \sigma^z_{\sigma(j)} |G_-\rangle =  2 \omega_{k=1}N!  \left [ \Pi_{j=1..N} (\frac{\omega_{F,j}}{2 \omega_{k=1}})  \right ]      \langle G_{+}|\Pi_{j=1..N} \sigma^z_{j} |G_-\rangle \\
   =  2 \omega_{k=1} N!  \left [ \Pi_{j=1..N} (\frac{\omega_{F,j}}{2 \omega_{k=1}})  \right ] e^{ \frac{-4 g^2}{ \sin(\frac{\pi}{2N})^2} \sum_{1\leq k_e \leq N_m} \frac{1}{k_e^3} } 
   \end{eqnarray}
The averaged degeneracy splitting $<\delta>$ and the standard deviation $\sigma$ over the disorder
configurations read:
\begin{eqnarray}
<\delta> \sim 2  \omega_{k=1} N! (\frac{\omega_{F}}{2 \omega_{k=1}})^N e^{ \frac{-4 g^2}{ \sin(\frac{\pi}{2N})^2} \sum_{1\leq k_e \leq N_m} \frac{1}{k_e^3} } \\ \nonumber
\sigma=\sqrt{<\delta^2>-<\delta>^2} \sim 2  \omega_{k=1} N! (\frac{\omega_{F}}{2 \omega_{k=1}})^N   e^{ \frac{-4 g^2}{ \sin(\frac{\pi}{2N})^2} \sum_{1\leq k_e \leq N_m} \frac{1}{k_e^3}} \sqrt{(1+ (\frac{\Delta}{\omega_F}) ^2)^N-1}  \sim  \sqrt{N} \frac{\Delta}{\omega_F} <\delta>
\end{eqnarray}
So, at resonance, and keeping only the dominant term,
\begin{equation}
\log{(\frac{<\delta>}{\omega_F})}= -\beta(N) g^2 \sim \frac{-4 g^2}{ \sin(\frac{\pi}{2N})^2} \sum_{1\leq k_e \leq N_m} \frac{1}{k_e^3} \\
\end{equation}
where $ 1.6 N^2 < \frac{4}{ \sin(\frac{\pi}{2N})^2}\sum_{1\leq k_e \leq N_m} \frac{1}{k_e^3} < 2.1 N^2 \,\,\,\,\forall N\geq 2 $, which  agrees with the numerical results .\\
The standard deviation $\sigma$ of the degeneracy splitting does depend on the normalized disorder amplitude $\Delta/\omega_F$, but it has the same exponential dependance  than the average splitting $\delta$. Hence, the effect of disorder can be made arbitrarily small when $g$ and/or $N$ are large enough.

\end{widetext}


\begin{thebibliography}{}
\bibitem{note_D}
In the case of the Hopfield model \cite{Hopfield},
the excitations are given by the eigenvalues of a matrix of the form
$$
\label{Hopfield}
  \mathcal{M}^{\mathrm{Hopfield}}_{k}=\left (
  \begin{array}{cccc} \omega_k + 2 D_k& -i \Omega_{k}& -2 D_k &-i \Omega_k\\ i \Omega_k  &  \omega_{12}
  &  -i \Omega_k& 0 \\ 2D_k & -i \Omega_k& -\omega_k -2 D_k&  -i \Omega_k \\  -i\Omega_k& 0 & i\Omega_k& - \omega_{12} \end{array}
  \right),
$$
where $\omega_{12}$ is the two-level transition frequency, $\Omega_k$
is the vacuum Rabi coupling (proportional to the electric dipole
matrix element) and $D_k$ is a term arising from the square vector
potential in the light-matter interaction. In the Hopfield
matrix\cite{Hopfield},  $D_k = \Omega_k^2/\omega_{12}$ and $Det(
\mathcal{M}^{\mathrm{Hopfield}}_{k}) = (\omega_k \omega_{12})^2
\neq 0$ and no quantum critical point exists, even if the coupling
can be ultrastrong\cite{Ciuti_vacuum,Ciuti_PRA}.
\end{thebibliography}
\end{document}